\documentclass[manuscript, screen, sigconf]{acmart}

\AtBeginDocument{%
  }

\usepackage{subcaption}
\usepackage{physics}
\usepackage{url}
\usepackage{makecell}
\usepackage[acronym]{glossaries}

\newacronym{qpd}{QPD}{quasi-probabilistic decomposition}
\newacronym{qps}{QPS}{quasi-probabilistic simulation}

\begin{document}

\title{Understanding the Scalability of Circuit Cutting Techniques for Practical Quantum Applications}
\author{Songqinghao Yang}
\affiliation{
  \institution{Cavendish Laboratory, Department of Physics,\\ University of Cambridge}
  \country{United Kingdom}
}
\email{sqhy2@cam.ac.uk}

\author{Prakash Murali}
\affiliation{
  \institution{Department of Computer Science and Technology, University of Cambridge}
  \country{United Kingdom}
}
\email{pm830@cam.ac.uk}


\begin{abstract}

Circuit cutting allows quantum circuits larger than the available hardware to be executed. Cutting techniques split circuits into smaller subcircuits, run them on the hardware, and recombine results through classical post-processing. Circuit cutting techniques have been extensively researched over the last five years and it been adopted by major quantum hardware vendors as part of their scaling roadmaps. We examine whether current circuit cutting techniques are practical for orchestrating executions on fault-tolerant quantum computers. We conduct a resource estimation-based benchmarking of important quantum applications and different types of circuit cutting techniques. Our applications include practically relevant algorithms, such as Hamiltonian simulation, kernels such as quantum Fourier transform and more. To cut these applications, we use IBM's Qiskit cutting tool. We estimate resources for subcircuits using Microsoft's Azure Quantum Resource Estimator and develop models to determine the qubit, quantum and classical runtime needs of circuit cutting. We demonstrate that while circuit cutting works for small-scale systems, the exponential growth of the quantum runtime and the classical post-processing overhead as the qubit count increases renders it impractical for larger quantum systems with current implementation strategies. As we transition from noisy quantum hardware to fault-tolerance, our work provides important guidance for the design of quantum software and runtime systems. 
 
\end{abstract}

\maketitle
\pagestyle{plain}
\section{Introduction}
\begin{figure}[t]
    \centering
    \includegraphics[width=1\linewidth]{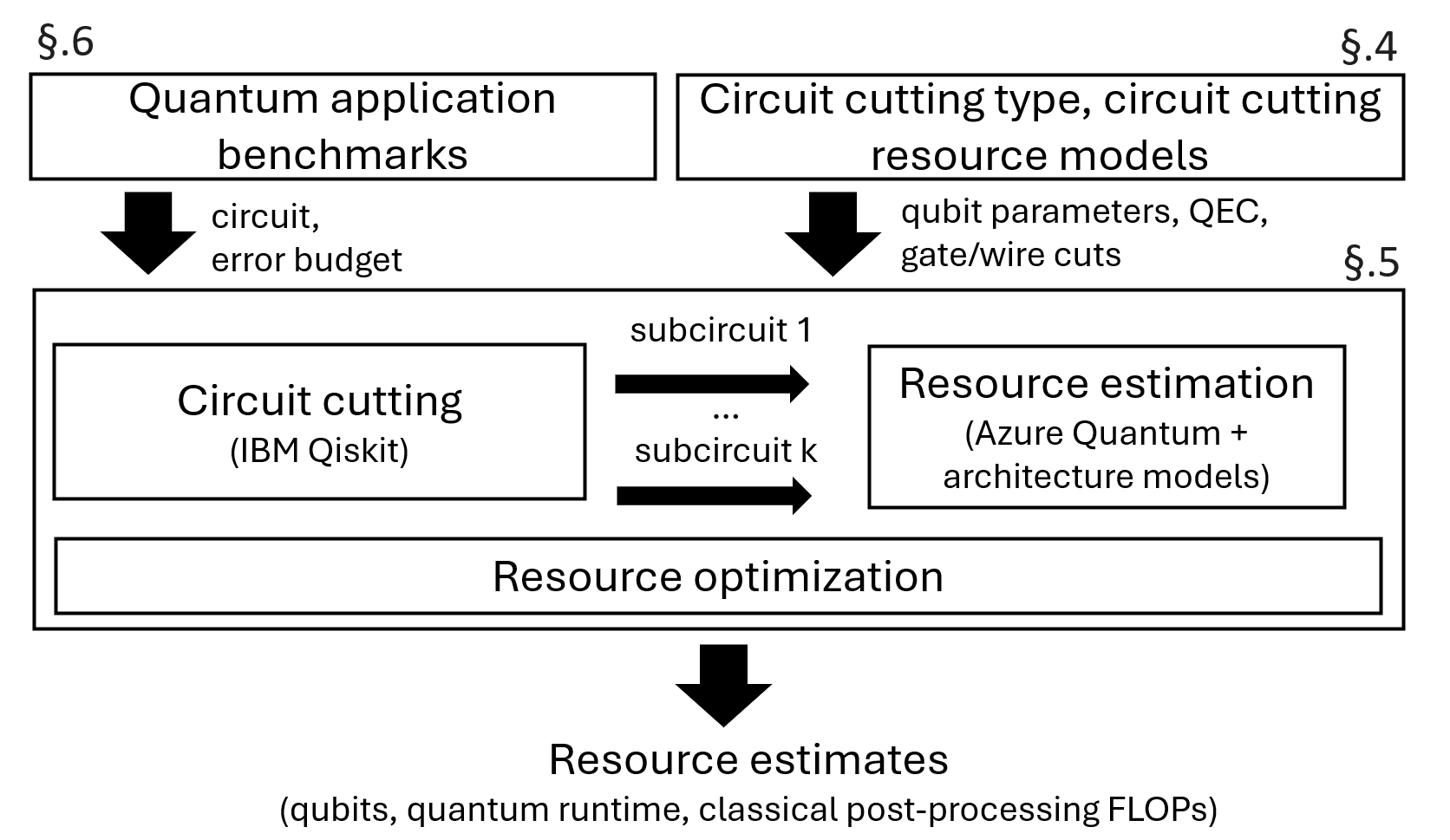}
    \caption{Our evaluation framework. Inputs are 1) a set of practically important quantum applications and desired error budget and 2) circuit cutting type and models for its resource usage in terms of number of quantum subcircuits, number of samples required and classical reconstruction overheads. Applications are cut and resources are estimated and optimized for subcircuits using models for a fault-tolerant QC system. Outputs include the number of cuts, physical qubits, and the quantum and classical runtime required for the application.}
    \label{fig:intro}
\end{figure}
Quantum computing (QC) holds transformative potential across numerous fields, including chemistry, material science and cryptography~\cite{bennett1992experimental, babbush2018encoding}. Current quantum hardware includes systems with 50 to 1000 qubits, depending on the physical qubit technology. These systems are popularly referred to as Noisy Intermediate Scale Quantum (NISQ) to indicate their restricted qubit count and limited gate quality~\cite{barral2024review}. In the last two years, QC has started transitioning from NISQ to fault-tolerance, with demonstrations of systems with a few \emph{logical} qubits that make use of Quantum Error Correction (QEC).

To achieve practical quantum advantage, that is, a speedup over classical computing on a useful problem, we require systems with several thousand logical qubits, potentially with a few million physical qubits~\cite{babbush2018encoding, lee2021even}. To progress towards this vision, it is important to understand which qubit technologies, software techniques and architectures are likely to scale up to meet the computational needs of practically useful applications and which techniques have scaling limitations. In this spirit, this paper examines the scalability of circuit cutting and knitting techniques for the upcoming fault-tolerant quantum era.  

Circuit cutting is a popular technique to execute QC algorithms that use more qubits than available in quantum hardware. Circuit cutting partitions large quantum circuits into smaller, manageable subcircuits that are executed on a device. The results from these executions are classically post-processed to reconstruct the result of the original circuit~\cite{peng2020simulating}. Circuit cutting has been extensively studied theoretically to explore different types of circuit partitions and optimizations for reducing its classical compute overheads~\cite{piveteau2022circuit, piveteau2022quasiprobability, Lowe2023fastcutting, Brenner2023optimal, pednault2023alternative}. It has also been experimentally validated, with demonstrations of QC circuit executions that go beyond small device sizes, improvements in application fidelity and several tools are available to optimize cut locations and orchestrate executions~\cite{tornow2024quantum, tang2021cutqc, Tang2022scaleqc}. Circuit cutting is seen as a promising technique to distribute quantum executions across multiple quantum processors in a manner akin to distributed computations in classical computing. Major hardware players feature circuit cutting techniques as part of their roadmap to achieve practical QC advantage~\cite{IBM, piveteau2022circuit, dell, vazquez2024scaling}.  

In this paper, we ask whether circuit cutting can be used to efficiently orchestrate executions of practically important quantum applications. In essence, we ask whether circuit cutting should be part of the long-term QC stack and whether components in the stack should be designed considering its classical post-processing or communication needs. This is challenging because of a number of factors. First, circuit cutting's resource requirements are closely tied to the structure of the circuits that we wish to partition and vary significantly across problem instances.
Second, there are a number of variants of the technique with different classical and quantum overheads, including the basic qubit and gate cuts~\cite{peng2020simulating, mitarai2021constructing, piveteau2022quasiprobability} and more complex types of cuts that depend on the gate set and communication assumptions~\cite{ufrecht2024optimal, Brenner2023optimal,piveteau2022circuit}. Third, practical resource requirements are influenced by the full stack, from the desired fidelity of the application, down to the choice of QEC methods and quality of the physical qubits~\cite{beverland2022assessing}. 

We conduct a comprehensive benchmark-driven resource analysis of circuit cutting techniques. Figure \ref{fig:intro} illustrates our approach.
To address the challenges above, we develop a benchmark suite that consists of a diverse set of \emph{practical-scale} quantum circuits. This is a crucial choice, since in applications like Hamiltonian simulation, only circuits that meet certain depth and connectivity thresholds can serve as meaningful physical models and offer practical QC use cases~\cite{vazquez2024scaling}.

To navigate the intricacies of different cutting techniques, we combine the Qiskit addon circuit cutting framework and the Azure Quantum Resource Estimator to evaluate resource estimates (such as physical qubit counts, runtime) for full applications orchestrated with circuit cutting. In contrast, existing works only provide overhead estimates at the level of individual gates. Further, we develop simple optimization techniques that reduce resource requirements through careful use of distillation factories and resource partitioning across subcircuits. Finally, we augment our benchmarking study with a theoretical analysis of circuit cutting requirements of each benchmark and use this to validate our experimental findings. 

Our contributions are as follows:
\begin{itemize}
\item Circuit cutting reduces physical qubit requirements on fault-tolerant QC systems by an average 30\%. However, obtaining qubit reductions comes at a prohibitive cost in terms of the large number of cuts, the runtime required on the quantum computer and classical post-processing time. Therefore, circuit cutting is not scalable for most practically useful quantum algorithms.

\item Most QC applications do not admit the connectivity structures required for obtaining a small number of cuts, with typical cases requiring a linear increase in the number of cuts as problem size increases.  

\item The choice of cutting method and type of gate that is cut have a significant impact on the quantum runtime and classical post-processing overhead. Certain benchmarks such as QAOA have manageable classical overhead, but large quantum runtime, while others such as QFT have manageable quantum overheads with exponential classical overheads. Therefore, we conclude that in spite of a wide variety of cutting techniques being developed~\cite{Brenner2023optimal, harada2023doubly, pednault2023alternative, brandhofer2023optimal}, resource requirements are extremely high across different cutting scenarios.

\end{itemize}

Our insights are valuable for the quantum community to adjust our understanding of long-term quantum stack and its architecture. While circuit cutting can still be used for small fault-tolerant executions, we demonstrate that long-term QC architecture should not be designed around circuit cutting capabilities.

\section{Background}
\begin{figure}[t]
    \centering
    \includegraphics[width=1\linewidth]{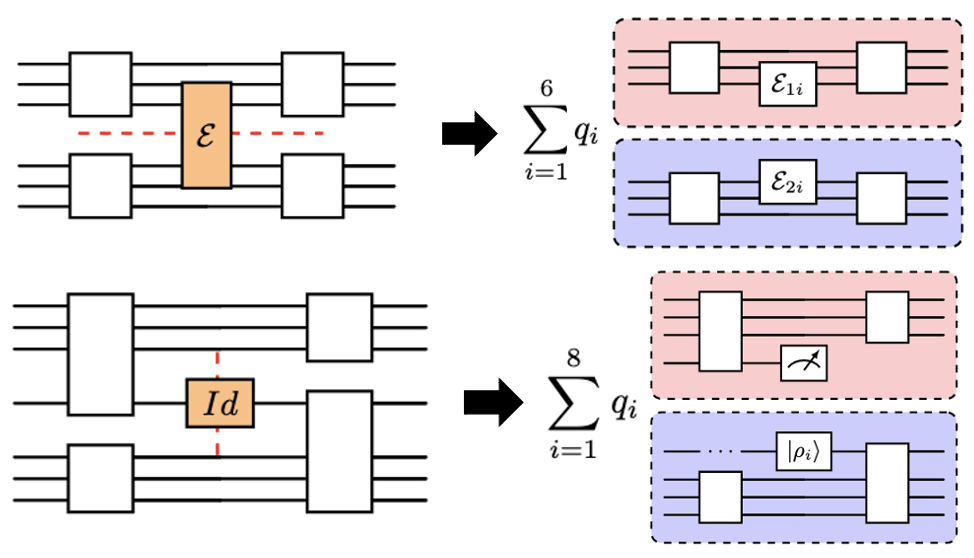}
\caption{Two schemes for cutting a quantum circuit: gate cutting on the top~\cite{mitarai2021constructing} and wire cutting at the bottom~\cite{peng2020simulating}. To obtain the final results, we need to run each subexperiment, resulting from the decomposition rules, with different coefficients $q_i$ and different local unitary gate (or prepare different initial state for the wire cut). The number of subexperiments for a gate cut is six while for one wire cut it is eight.}
    \label{cutting-schemes}
\end{figure}

In this section, we introduce foundational concepts and terms relevant to the implementation of quantum circuit cutting and resource estimation in the fault-tolerant setting.

\subsection{Circuit Cutting Basics}

Circuit cutting involves three distinct phases:
\begin{itemize}
    \item \textbf{Cut}: In this phase, a large quantum circuit is divided into smaller, manageable subcircuits by strategically introducing ``cuts'' at specific points. This is typically achieved by inserting operations such as mid-circuit measurement or other unitaries that allow the system to be partitioned while preserving essential quantum information.
    \item \textbf{Run}: The subcircuits are then executed independently on quantum hardware or simulators. The choice of the subcircuits to run will depends on two factors: the sampling number $N_s$ and the pool of subexperiments $C$, constructed from the subcricuits $c_i \in C = \{c_1 \dots c_{N_{\text{pool}}}\}$, with $N_{\text{pool}}$ being the total number of subexperiments in the pool. More explanation on this concept will come in later sections. After cutting the circuit, one needs to uniformly sample $N_s$ subcircuits from the pool $C$ to execute and record the measurements.
    \item \textbf{Reconstruct}: The results from the subcircuits are combined mathematically with tensor networks as kronecker delta product to recreate the outcomes of the original circuit. This phase often involves classical post-processing to integrate the data and correct for any approximations introduced during the cutting phase. Typically, if there are $n$ cuts in a circuit with $N$ qubits, the result tensor would be in shape $4^n\times2^N$, which has $n$ legs with 4-dimension from decomposition and one leg with $2^N$-dimension, which is the output for the original circuit~\cite{ren2024hardware, Tang2022scaleqc}.
\end{itemize}

\subsection{Fault-tolerant Quantum Computing}

Fault-tolerant quantum computation (FTQC) is distinguished from NISQ devices by its use of QEC to reliably execute arbitrarily long computations, while NISQ systems operate without error correction, limiting their scalability and susceptibility to noise. By encoding logical qubits into entangled states of multiple physical qubits, QEC schemes can detect and correct errors without directly measuring the logical qubits. Despite the application of QEC, a logical error  $\epsilon_{\text{log}}$ occurs when an error bypasses the protective encoding and directly affects the logical qubit.

In addition, certain non-Clifford operations, essential for universal quantum computation, cannot be directly implemented fault-tolerantly. Distillation is used to generate high-fidelity \textit{magic states} or other specialized resources required for these operations. This process typically involves iterative procedures to purify a resource while sacrificing less-perfect copies, resulting in a small final deviation from the non-Clifford operation $\epsilon_{\text{dis}}$.

Another important process is the construction of synthesis gates, especially for rotational gates. Since universal gate sets in FTQC (e.g., Clifford+T) cannot directly implement arbitrary rotations, synthesis allows these rotations to be approximated to the required error, $\epsilon_{\text{syn}}$, by decomposing them into a sequence of gates from the fault-tolerant set.

Considering all above points, plus the intrinsic error that comes with the quantum algorithm itself, $\epsilon_{\text{alg}}$, we use the term \emph{overall error budget} to define the acceptable level of error, computed as:   
\begin{equation}
    \epsilon = \epsilon_{\text{log}} + \epsilon_{\text{dis}} + \epsilon_{\text{syn}} + \epsilon_{\text{alg}}
    \label{errorbudget}
\end{equation}
in the final results of a quantum computation. 

In the quest for fault-tolerant, error-corrected quantum computers—still potentially years away—accurate resource estimation has become a crucial factor in understanding the development of quantum algorithms and assessing trade-offs between different architectural designs. 

\section{Resource Requirements and Tradeoffs}
\subsection{Structure of the application} 

The patterns of two-qubit gates in an application have a significant influence on the resource requirements of circuit cutting. If an application's qubits can be grouped into a set of clusters with gates being predominantly performed within the cluster, it is relatively easy to partition the circuit using a small number of cuts. On the other hand, when gates are more uniformly distributed across the qubits, it is hard to find partitions that use only a few wire or gate cuts. Since the quantum and classical runtime overheads of circuit cutting increase exponentially with the number of cuts, it is crucial that the algorithm structure is highly-clustered. For example, consider a 2D Ising model application - here program qubits are arranged in a 2D grid and have only nearest-neighbor logical gates among them; therefore, circuit cutting is unlikely to succeed in this application. 

However, it is not easy to determine the suitability of circuit cutting from only inspecting the connectivity pattern of an application. For applications such as Hamiltonian simulation a common approach is to use Trotterization to implement the quantum circuit. Here, the circuit consists of a set of steps, with each step having the same pattern of gates. Usually, a large number of number of steps required for problem sizes and accuracy levels that are relevant for practical quantum advantage. Therefore, even when individual steps exhibit good clustering structure, the sheer number of repetitions can lead to a large number of cuts and subcircuits. In addition, the native gate set that the circuit transpiles into affects the overhead for the cutting. For example, to express the Trotterization with $R_{zz}$ gate instead of the CX gate will have a significant reduction in the overhead.

While several works have studied the feasibility of circuit cutting for NISQ applications, our work aims to explore whether workloads that are important for practical quantum advantage exhibit such clustered structure. 

\subsection{Types of cutting techniques}
There are several ways of cutting a quantum circuit into subcircuits, with varying resource requirements. Circuit cutting can be primarily classified into two methods: `wire cutting' and `gate cutting'~\cite{harada2023doubly, ufrecht2023cutting} and are illustrated in Figure \ref{cutting-schemes}. As the name suggests, qubits are cut in the former and two-qubit gates are cut in the latter.  The total samples required for reconstruction with error $\varepsilon_{\text{rct}}$ are $\mathcal{O}(16^{n_{\text{wire}}} / \varepsilon_{\text{rct}}^2)$ and $\mathcal{O}(9^{n_{\text{gate}}} / \varepsilon_{\text{rct}}^2)$, respectively, for the wire cut and gate cut (CX)~\cite{peng2020simulating, mitarai2021constructing}. For gate cuts, the overhead is dependent on the type of gate that is cut and even the rotation angles that may parameterize it. Table \ref{gate_table} summarizes these overheads~\cite{qiskit-addon-cutting}. Therefore, two applications with the same patterns of two-qubit gates may have entirely different circuit cutting overheads owing to differences in the exact type of two-qubit gates they use. 

\begin{table}[t]
\caption{The table above provides the sampling overhead factor for various two-qubit instructions, provided that only a single instruction is cut. Note that gates like CYGate, RYYGate, and CRZGate have the same overhead as their X counterparts. For simplicity, we do not list them all here. Collected from~\cite{qiskit-addon-cutting}.} 
\centering
\begin{tabular}{|c|c|}
\hline
Instruction & Sampling overhead factor \\
\hline
CSGate, CSXGate&$3+2\sqrt{2} \approx 5.828$\\
\hline
CXGate, CHGate&$3^2=9$\\
\hline
iSwapGate, SwapGate&$7^2=49$\\
\hline
RXXGate, RZXGate&$\left[1 + 2 \left|\sin(\theta)\right| \right]^2$\\
\hline
CRXGate, CPhaseGate&$\left[1 + 2 \left|\sin(\theta/2)\right| \right]^2$\\
\hline
\end{tabular}
\label{gate_table}
\end{table}

\subsection{Full-stack resource influence and tradeoffs}

Circuit cutting resource requirements are also heavily influenced by application's error requirements and the architecture of the quantum computer. As the required error for an application decreases, the resource requirements tend to increase, both in terms of physical qubit counts and runtime. Such behavior is typically understood for single quantum circuits, but when circuit cutting is used, we have several subcircuits each with its own error. If we have k subcircuits, a naive approach is to run each subcircuits with error $\epsilon/k$. However, this may be suboptimal in terms of resource requirements since subcircuits may have different qubit counts, gate depths and magic state requirements. Our explores explores this optimization opportunity and its impact on the practicality of circuit cutting. 

The physical resource requirements with cutting depend on the type of error correction code and distillation factories. We assume surface codes for this work, a standard choice for QEC on superconducting qubits and other solid state platforms~\cite{kitaev2003fault, fowler2012surface}. For distillation, we consider two aspects. First, the type of distillation circuit that is used has an impact on the error of magic states, which in turn influences the error of the circuit that uses it and its resource requirements. Second, the number of distillation factories influences qubits and runtime: more factories mean that subcircuits run faster, but at the expenses of higher physical qubit counts.

In summary, we ask: \emph{``What are the qubit and runtime demands when circuit cutting is employed to run applications? Do practical-scale quantum applications have the necessary clustering structure to benefit from circuit cutting? How does the instruction mix of the application and cutting method influence the scalability of circuit cutting? How much do other factors such as the subcircuit errors and distillation choices affect circuit cutting performance?''}

\subsection{Overview of our approach}
To answer the design questions, we develop the toolflow shown in Figure \ref{fig:intro}. Using a benchmark that consists of practically-relevant quantum applications and a model for a surface code-based quantum computer, we conduct resource estimation studies to determine the impact of circuit cutting on application qubit counts and runtime needs. 

To enable this study, we first develop models to quantify the resource needs of circuit cutting. To apply these models for applications, we use IBM's Qiskit circuit cutting tools to find optimized cuts. We used Microsoft's Azure Quantum Resource Estimator to determine resources for executing the subcircuits from the cut circuit. We also develop simple optimizations to reduce resource needs through error and distillation factory optimizations. 

\section{Modelling Circuit Cutting Executions}
Given a quantum circuit and a cutting method such as wire or gate cutting, we develop models that capture the logical qubit requirements, number of subcircuits, quantum runtime and classical post-processing overhead. These models are based on known theoretical foundations of circuit cutting.

Circuit cutting uses the \gls*{qpd} of a quantum channel, which evolves \(\rho_0\) to \(\rho\) as \(\mathcal{E}(\rho_0)\). Then using the Kraus decomposition~\cite{kraus1971general}, \(\mathcal{E}(\rho) = \sum_{j=1}^m E_j \rho E_j^\dagger\), this representation allows flexibility in choosing local Kraus operators \(E_j\). For a bipartite system \(\rho = \rho^{(1)} \otimes \rho^{(2)}\), a channel \(\mathcal{E}\) acting non-locally can be decomposed into local operations:
\begin{equation}
\mathcal{E}(\rho) = \sum_{i=1}^m q_i \mathcal{E}_i^{(1)}(\rho^{(1)}) \otimes \mathcal{E}_i^{(2)}(\rho^{(2)}),
\end{equation}
where \(q_i\) are quasi-probabilities (\(\sum q_i = 1\)), allowing the separation of qubits for circuit cutting.

The effectiveness of circuit cutting techniques is significantly limited by the exponential increase in runtime required to stitch together the results from subcircuits. Circuit cutting entails classically monitoring all quantum degrees of freedom at each cut point, which naturally leads to an exponential growth in computational demand as the quantum state size expands. This exponential overhead in circuit execution remains a fixed cost of the \gls*{qpd} decomposition process, regardless of the goal of the circuit cutting application~\cite{Marshall2023bounds}. 

Specifically, we categorize the overhead of circuit cutting into 1) the \textit{runtime sampling overhead} and (2) the \textit{classical recombination post-processing overhead}. 

\noindent \textbf{Runtime sampling overhead}: This term refers to the total runtime required on the quantum computer when circuit cutting is used to run an application. This is dependent on two factors. First, when a circuit is cut, the number and type of the cuts determine the number of degrees of freedom that must be tracked at each cut, leading to a set of subcircuits. From this set, a subset is sampled based on the reconstruction error required in the expectation value of the observable of interest. Second, each subcircuit that is sampled is run on the quantum computer and the architecture of the device influences its runtime and qubit needs.

It means our total quantum runtime equals to the number of samples times the quantum runtime of each sample. Formally, the first term can be expressed as:
\begin{equation}
    \label{sampling_overhead}
    \mbox{Sampling overhead}:=\gamma(\mathcal{E})^2,~~ 
    \gamma(\mathcal{E}):=\sum_{i=1}^{m}|q_{i}|.
\end{equation}
In practice, the subcircuits are executed with the total number of samples $N_s$ to achieve a desired error $\varepsilon_{\text{rct}}$:
\begin{align}\label{eq:old_samplingcost}
     N_s=O\left({\gamma^{2}}\times\frac{1}{ {\varepsilon_{\text{rct}}^2}}\right),
\end{align}
which comes from the Hoeffding’s inequality of the Monte-Carlo sampling. 
Furthermore, if we apply the cutting at $n$ places in a quantum circuit, the total number of measurements with the error $\varepsilon_{\text{rct}}$ results in $O(\gamma^{2n}/\varepsilon_{\text{rct}}^2)$. The total samples required for reconstruction with error $\varepsilon_{\text{rct}}$ were first shown in~\cite{peng2020simulating, mitarai2021constructing} to be $\mathcal{O}(9^{n_{\text{gate}}} / \varepsilon_{\text{rct}}^2)$ and $\mathcal{O}(16^{n_{\text{wire}}} / \varepsilon_{\text{rct}}^2)$, respectively, for the gate cut (CX) and wire cut. 

More specifically, our task is to estimate the expectation $\mathbb{E}[f(y)]$ for a random bitstring $y\in\{0,1\}^N$ sampled from the original circuit~\cite{bravyi2016trading}. The input state $\rho$ is initialized in $\ket{0}\bra{0}^{\otimes N}$ and the observable is calculated from some output function $f:\{0,1\}^N\to [1,-1]$. At each space-like cut, we get 6 different sets of single-qubit operations, so $n_{\text{gate}}$ cuts induce $6^{n_{\text{gate}}}$ terms. Likewise, $n_{\text{wire}}$ time-like cuts induce $8^{n_{\text{wire}}}$ terms. We call each of these partition-induced circuit instances a subexperiment. The total number of subexperiments are then $6^{n_{\text{gate}}}8^{n_{\text{wire}}}$. we can then take a Monte-Carlo approach to estimate the sum, that is, we randomly choose $N_s$ circuits with a uniform distribution in the pool of the subexperiments to run and average them. 

We assume a space-efficient model for the subcircuit executions. That is, when we have a collection of subcircuits to execute, we execute them one after another on the QC system. This leads to the minimum number of logical and physical qubits, at the expense of runtime. This approach has two benefits. One, while it increases the overall runtime due to the sequential execution of subcircuits, it significantly reduces the total qubit resources necessary for synthesis and distillation factories. Two, this model allows us to determine the best case qubit reductions possible from circuit cutting.  Another approach would be to consider a time-efficient model where subcircuits are run in parallel on a large number of small QC systems, leading to high qubit usage and reduced runtime. The maximum benefits from such an execution are at best a linear speedup in the quantum runtime, but classical post-processing costs will not change. 

In the space-efficient model, with each subexperiment having $Q_c$ logical qubits, the number of logical qubits required is 
\begin{align}\label{qubitneeds}
 Q_{\text{max}}=\max\limits_{c \in \{\text{subexperiments}\}} Q_c.
\end{align}
The quantum runtime is upper bounded by
\begin{align}\label{samplingcost}
     T_{\text{max}}=\max\limits_{c \in \{\text{subexperiments}\}} T_c \times N_s.
\end{align}
where $T_c$ is the quantum runtime for the subexperiment $c$. We determine $T_c$ using resource estimation.

\noindent \textbf{Classical recombination post-processing overhead}: This term refers to the amount of classical post processing required to reconstruct results after circuit cutting. The process of post-processing, the contraction of two subcircuits, is equivalent to a pairwise tensor contraction. The memory requirement for such post-processing arises from two main components~\cite{Tang2022scaleqc}:
\begin{enumerate}
    \item Storage of input subcircuit tensors: The tensor for each subcircuit has dimensions \(4^{n} \times 2^{N}\) where $n$ is the number of cuts and $N$ is the total number of qubits in the original circuit.
    \item Storage of intermediate tensors: These arise from products computed at each contraction step.
\end{enumerate}

The compute cost is determined by the number of floating-point multiplications (FLOPs) in each contraction. In~\cite{Tang2022scaleqc} the authors has reduced the number of multiplications one needs to do by contracting the subcircuit sequentially instead of doing the contraction at one go, removing the redundancy of those unaffected subcircuits for wire cuts. Ref.~\cite{ren2024hardware} explores similar tensor contraction optimizations for gate cuts.  However, both the memory and compute requirements scale with the number of cuts per subcircuit in each contraction step, $\text{FLOPs required} \sim O\left(4^n\right)$.

\section{Resource Estimation and Optimization for Circuit Cutting}
The input to our toolflow in Figure \ref{fig:intro} is the circuit for a quantum application and the required error $\epsilon$. We cut this circuit into a number of subcircuits. Logical circuit cutting requirements (such as $Q_{max}$, $T_{max}$) that were described in the last section are translated into physical resource estimates using a resource estimator and resource optimization techniques.

\subsection{Circuit Cutting Method}
 We require a method that can partition a circuit into smaller subcircuits. Several past works have explored cutting techniques that reduce the sampling overhead. Optimization based techniques such as CutQC~\cite{tang2021cutqc} do not scale to the problem sizes that we use, therefore we require heuristics. We use IBM's Qiskit circuit cutting addon tool \cite{qiskit-addon-cutting}. This tool implements gate and wire cuts, with Dijkstra's best-first search. This tool allows us to specify the \emph{maximum number of qubits per subcircuit}, to control the number and sizes of subcircuits. For example, in Figure 3, a 6 qubit circuit is partitioned into two subcircuits with a maximum of 4 qubits per subcircuit. Although a heuristic, our theoretical analysis of cut counts of applications in Section \ref{theoretical} shows that this tool produces cuts that are near-optimal, justifying its use for benchmarking.
 
\subsection{Resource Estimation of Subcircuits}

For each subcircuit that is generated from the cutting, we wish to estimate the number of physical qubits required to run it and the expected runtime on a quantum processor. To accomplish this, we require 1) a resource estimation tool and models for a fault-tolerant quantum stack 2) the error budget with which each subcircuit must be executed. 
 
Several resource estimation tools have been developed to provide estimates with different levels of granularity. For example, TFermion~\cite{casares2022tfermion} offers estimates for a broad range of quantum chemistry algorithms but sometimes overestimates resources due to its strict error bounds. At the same time, OpenFermion~\cite{mcclean2020openfermion} focuses on specific quantum chemistry methods and provides some insight into surface code overheads. Similarly, Google's Qualtran~\cite{harrigan2024expressing} and Zapata's BenchQ~\cite{benchq} provide advanced frameworks for estimating resources, each specializing in different quantum architectures and applications. Among these tools, we use Microsoft's Azure Quantum Resource Estimator (AQRE)~\cite{beverland2022assessing} since it works with real algorithm implementations, models the full-stack of a fault-tolerant system and provides state-of-the-art estimates. 

For an input circuit, AQRE counts the number of logical qubits, logical operations, depth and determines the required number of magic states. Based on the required error of the circuit, it computes the desired logical error rates. We model a superconducting qubit system with surface code error correction, using the physical qubit and error correction models provided in AQRE. AQRE converts the logical estimates into physical estimates by applying the resource models for error correction and distillation. It outputs the number of physical qubits and required quantum runtime.

To determine the total resources needed for a collection of subcircuits, we assume the space-efficient execution model described in the previous section. Therefore, the number of physical qubits is determined as the maximum number of physical qubits needed for any subcircuit. That is, the total qubit count is driven by the peak qubit demand of individual subcircuits rather than their cumulative requirements. The total runtime is obtained by equation \ref{samplingcost}, where $T_c$ is the physical runtime estimated from AQRE.

\subsection{Error budget and Distillation Optimizations}
\textbf{Error budget:} To set the error budget with which each subcircuit must execute, a naive option is set it as $\epsilon/k$ where k is the number of subcircuits. However, when the maximum number of qubits per circuit is high, we may have subcircuits where qubit counts are highly imbalanced. For example, a 100 qubit circuit could be cut into two subcircuits with 90 and 10 qubits each. In this setting, the error budget can be partitioned such that the 90 qubit circuit can use more error budget. This allows the 90 qubit circuit to run with a lower QEC code distance, since code distance is inversely proportional to the available error margin. A surface code logical qubit with distance $d$ uses $2d^2$ physical qubits. Therefore, this optimization reduces physical qubit usage, further favoring circuit cutting resource estimates. This optimization also reduces runtime of the largest subcircuits since the logical cycle time of the surface code is linearly proportional to the code distance. We implement this optimization by proportionally dividing the error budget across subcircuits, in the ratio of their qubit counts. Across our benchmarks, we observed up to 0.5\% to 10\% reductions in the physical qubit counts compared to equal error budget splitting.

\noindent\textbf{Distillation factories:} The number of distillation factories determines the rate at which magic states are produced and supplied for running non-Clifford gates. Higher values accelerate magic state production, thereby reducing runtime but increasing the demand for physical qubits. AQRE optimizes runtime, therefore it uses a large number of distillation factories to achieve minimum runtime. Experimenting with different values for number of factories, from a baseline of one factory, we observed a gradual runtime reduction of five fold from 167\% down to 33\% as the factories increased, compared to the baseline runtime. (Here we only consider a single instance's execution time, $T_c$ of each of the subcircuits $c$ compared to the original circuit's run time, $\frac{\sum T_c}{T_{\text{basline}}}$) However, we also found that the physical qubit requirements almost doubled at a T factory value of count for Hamiltonian simulation, increasing from 56\% to 87\%, compared to the baseline. Since this does not favor circuit cutting, we use a small number of factories in our experiments, from one to five, depending on the benchmark. Further runtime improvements from increasing distillation are at best one order of magnitude; our experiments show that this is not sufficient to make circuit cutting competitive for large benchmarks.

\begin{figure*}
    \centering
    \includegraphics[width=1\linewidth]{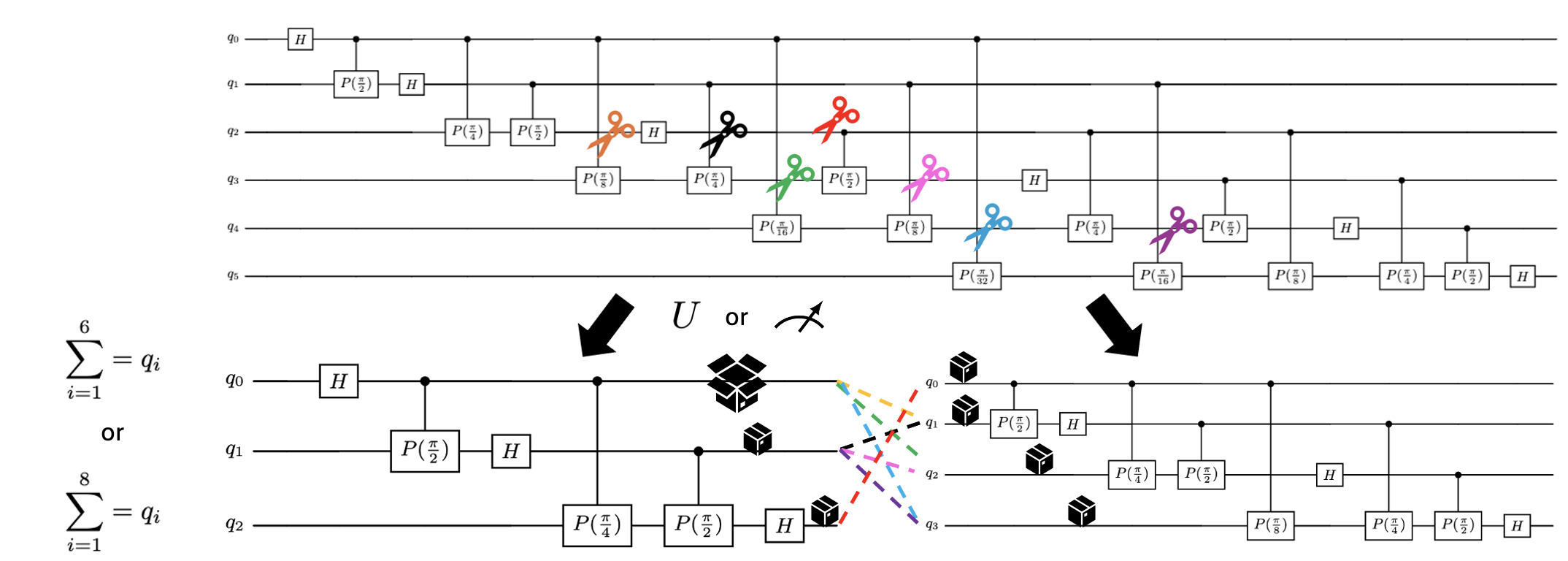}
    \caption{Cutting locations for a 6-qubit QFT circuit with minimum number of cuts to separate it into two subcircuits with 3 and 4 qubits respectively. The dotted connecting lines show how the subcircuits were originally connected with different colors corresponding to different scissors/cuts. The box contains the unitary or (mid-circuit) measurement operation that varies according to the type of the cuts.}
    \label{qftexample}
\end{figure*}

\noindent \textbf{Example:} We now illustrate the resource calculations using the 6-qubit Quantum Fourier Transformation (QFT) circuit, as shown in Fig~\ref{qftexample}. This circuit is partitioned into two subcircuits (3 and 4 qubits). From Table~\ref{gate_table}, we estimate the runtime sampling overhead as $\gamma^2_{\text{gate1}} \cdots \gamma^2_{\text{gate6}} \cdot \gamma^2_{\text{wire1}} \approx 460$, accounting for six gate cuts and one wire cut. To maintain an error margin of 1\%, the total number of samples required is $N_s = \frac{460}{0.01^2} = 4.6 \times 10^6$,
chosen from a pool of \(6^6 \times 8^1 = 373,248\) possible subexperiments. 

Using AQRE, the original circuit’s quantum runtime is approximately 0.018 seconds for a superconducting qubit system (see assumptions in experimental setup), with a total requirement of 11,320 physical qubits. By applying the error budget optimization, we allocate 80\% of the resources to the 4-qubit subcircuit and 20\% to the 3-qubit subcircuit. Consequently, the maximum runtime per subcircuit is reduced to approximately 0.0055 seconds, with a 17\% reduction in the number of physical qubits. In this example, the discrepancy of 3\% reduction, between the theoretical resource reduction and experimental reduction, compared to the baseline, comes from the resource allocation for the sub-components in the synthesis, distillation, algorithmic and logical error for each subcircuit. Therefore, to keep the error within the 1\% threshold, the total quantum runtime upper bound is estimated as $0.0055 \times 4.6 \times 10^6 \approx 7 \text{ hours}$. Additionally, the classical post-processing recombination requires a computational overhead of $4^7 \times 2^6 = 1,048,576$ multiplications.

\section{Experimental Setup}
\begin{table}
\caption{Summary of the benchmarks we used in this work. The middle column illustrates the logical qubit requirement for each algorithm depending on the problem size. Logical depth here means the circuit depth that excludes resources for T factories, in log scale.} 
\centering
\begin{tabular}{|c|c|c|}
\hline
Algorithm (problem size) & Logical qubits & Logical depth \\
\hline
Heisenberg $D=\{3,4,5,6\}$&$N = D\times D$&$10^{4.7},\cdots,10^{5.1}$ \\
\hline
Ising $D=\{3,4,5,6\}$&$N = D\times D$&$10^{4.2},\cdots,10^{5.2}$\\
\hline
Fermi-Hubbard $D=\{2,3,4\}$&$N = 2\times D\times D$&$10^{4.1},\cdots,10^{4.8}$\\
\hline
QFT $N=\{5,\cdots,60\}$&$N$&$10^{2.3},\cdots,10^{4.0}$\\
\hline
FABLE $2^N, N=\{2,3,4,5\}$&$N+1$&$10^{2.4},\cdots,10^{4.3}$\\
\hline
QPE $N=\{2,4,6,8\}$&$2\times N$&$10^{2.2},\cdots,10^{4.7}$\\
\hline
QAOA $N=\{10,\cdots,100\}$&$N$&$10^{2.9},\cdots,10^{3.5}$\\
\hline
Random $N=\{5,\cdots,20\}$&$N$&$10^{2.7},\cdots,10^{4.4}$\\
\hline
\end{tabular}
\label{logical_table}
\end{table}

\subsection{Benchmarks}
Our benchmark consists of set of applications that are relevant for practical use cases of QC systems. Hamiltonian simulation with Ising, Heisenberg, Fermi-Hubbard models are expected to be one of the first scientific applications where we see practical quantum advantage over classical computing~\cite{Daley2022, beverland2022assessing}. QFT is a core component of Shor's factoring algorithm~\cite{shor}. Quantum Phase Estimation and block encoding are important for quantum chemistry computations. Although quantum advantage prospects are not theoretically clear for QAOA, we include it in our benchmarking. While prior works have studied circuit cutting on some of these benchmarks, we choose benchmark instances that reflect practical computations. For example, for Hamiltonian simulation, we choose sufficient Trotter steps to allow highly accurate simulations, based on parameters from \cite{Daley2022, beverland2022assessing}. In contrast, \cite{ren2024hardware}, uses one or two Trotter steps, which is insufficient for practically useful results from the benchmark. In several cases (examples: QFT beyond 60 qubits, Ising models with 10x10 qubits with 80 Trotter steps), IBM's cutting tool fails to determine cuts because of large circuit size. In these cases, our benchmark includes the largest sizes we could cut. We will denote $N$ as the total number of qubits and $m$ as the maximum number of qubits allowed in a subcircuit.

\noindent \textbf{Hamiltonian Simulation:} The goal of Hamiltonian simulation is to approximate the time evolution of a quantum system under a given Hamiltonian. This is particularly important for studying the behaviour of quantum materials, chemistry, and particle physics. We choose the fourth-order Trotterization algorithm to keep the error small while maintaining feasibility.

- \textbf{Heisenberg Model}: This model describes interactions between spins in a lattice, accounting for all spin directions (x, y, z). It is widely used in quantum magnetism and includes both nearest-neighbor interactions and external magnetic fields.

- \textbf{Ising Model}: A simpler spin model compared to the Heisenberg model, it focuses on interactions along specific spin directions (z for coupling, x for the external field). It is often applied to systems where spins primarily align or oppose along these directions.

- \textbf{Fermi-Hubbard Model}: Unlike the spin-focused models above, this model describes fermions (particles with half-integer spin) on a lattice. It includes terms for particle hopping between sites and on-site interactions, making it crucial for studying strongly correlated electron systems, such as in materials science.

\noindent \textbf{Quantum Fourier Transformation (QFT):}
At a high level, the QFT circuit begins with a Hadamard gate applied to the first qubit, followed by controlled phase gates that entangle it with the remaining qubits. This process is then repeated recursively for the remaining qubits, excluding the one already processed. Finally, the qubits are swapped to reverse their order, ensuring the correct output state.

\noindent \textbf{Fast Approximate Block Encoding (FABLE):} The FABLE algorithm~\cite{camps2022fable} approximates block encodings of large non-unitary matrices. For a matrix \(A\) of size \(2^N \times 2^N\), block encoding reduces it to a unitary form that acts on \(N+1\) qubits, allowing quantum algorithms to handle non-unitary operations efficiently. It is widely used in various applications, such as qubitization and quantum singular value transformation. 

\begin{figure*}
\centering
\includegraphics[width=2.1\columnwidth]{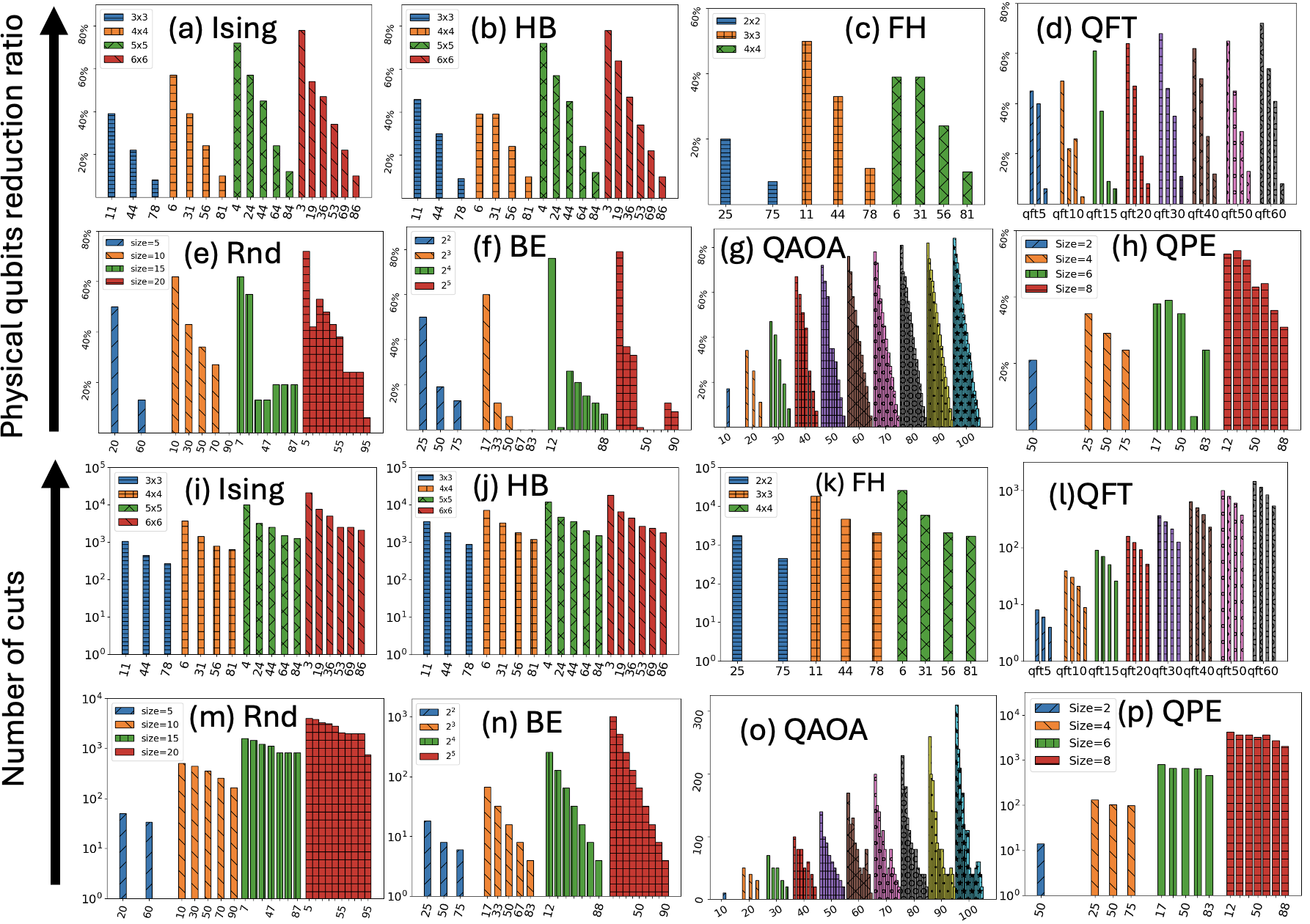}
    \caption{Data plots for our experiments. The first two rows summarizes the results for the physical qubit reduction and total number of cuts is summarized in row 3 and row 4 for each of the benchmarks. In the inset on the left corner in each plot, the total number of logical qubits are displayed. The data gaps occur in experiments where the number of physical qubits does not show a reduction, indicating a failure of the cutting approach in these cases. Accompanying axes and data interpretations are in the text. Note some of the x-axis labels are omitted for clarity.}
    \label{algo}
\end{figure*}

\noindent \textbf{Quantum Phase Estimation (QPE):} Quantum Phase Estimation (QPE) is a fundamental algorithm in quantum computing used to estimate a unitary operator's eigenvalues. The primary circuit structure of QPE consists of two central registers: a control register and a target register. The control register contains qubits initialized in a uniform superposition created by applying Hadamard gates to each qubit. The target register holds the eigenstate of the unitary operator whose phase (eigenvalue) we wish to estimate.

\noindent \textbf{Quantum Approximate Optimization Algorithm (QAOA):} QAOA is a hybrid quantum-classical algorithm used to solve combinatorial optimization problems. It alternates between applying a problem Hamiltonian and a mixing Hamiltonian. The depth of the circuit increases with the number of rounds \(p\). Here we have chosen $p=10$. Throughout the experiments, we just used a set of regular graphs for which the node number scales with the qubit number.

\noindent \textbf{Random unitary circuit:} Finally, we have created a random circuit with alternating layers of single-qubit unitaries and two-qubit unitaries. This model is benchmarked to demonstrate the infeasibility of circuit cutting when the density of multi-qubits gate/entanglement rises in the system. But we should note that this is not the most rigorous model as we have not properly introduced a way to measure the span of the entanglement across the orginal circuit comparing to some local entanglement. Future work could be studied, for example, to consider entanglement in the system as a clustering graph, to make the model more complete.

The logical requirements for each of above algorithms are summarized in the TABLE~\ref{logical_table}.

\begin{figure}
    \centering
    \includegraphics[width=1\linewidth]{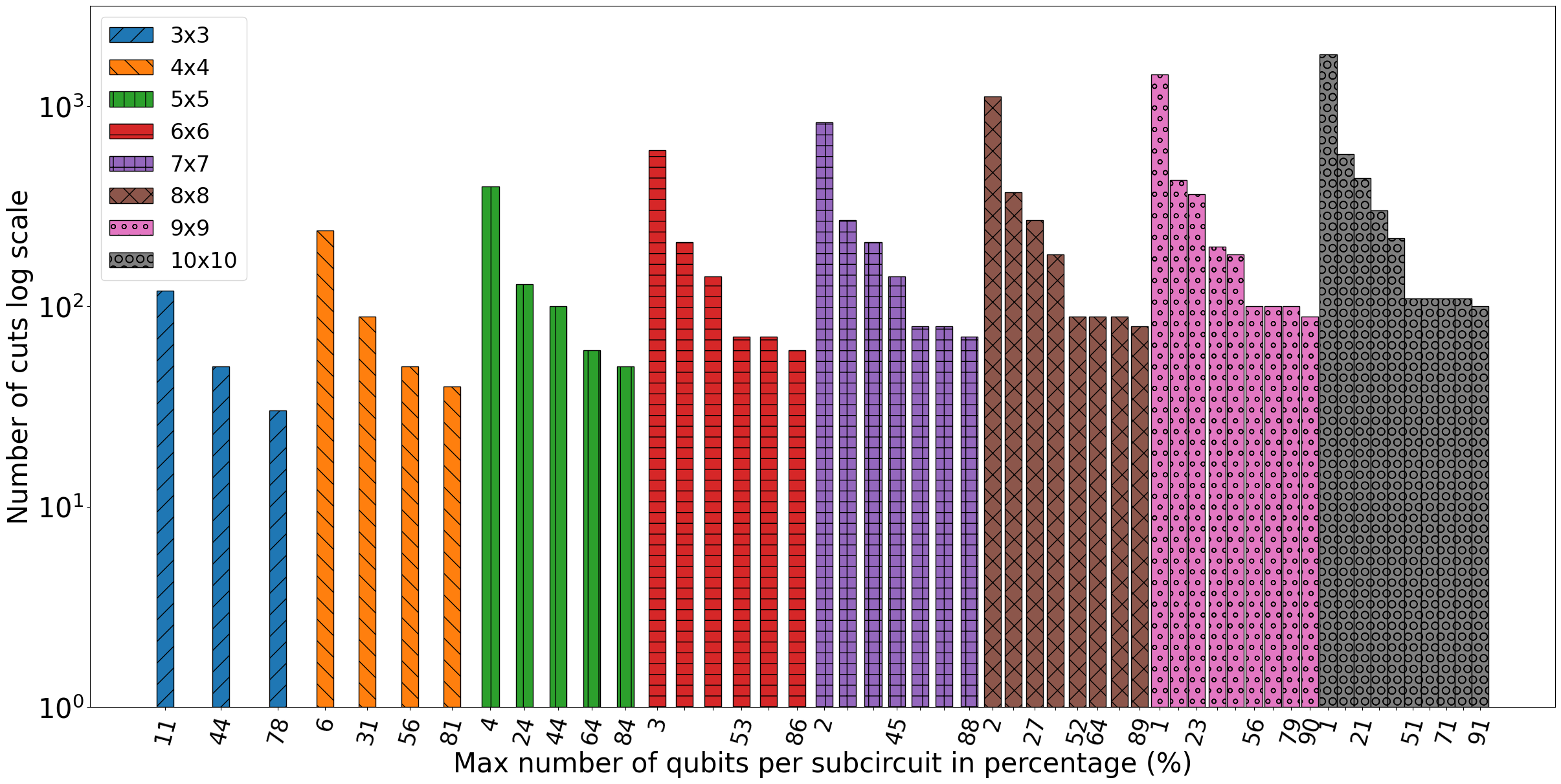}
    \caption{Single trotter step 2D Ising model.}
    \label{single}
\end{figure}

\subsection{Implementations and tools}
\textbf{Benchmark implementation and cutting:}
Our benchmarks were implemented in IBM Qiskit with Python (version 3.12.5). Circuit cutting and manipulation were performed using the Qiskit add-on \texttt{circuit-cutting} (version 0.9.0).

The circuits developed for Hamiltonian simulations utilize native two-qubit gate sets, including the $R_{zz}, R_{xx}, R_{yy}$ for entangling operations. Additionally, for the QFT circuits, controlled-phase gate is used, while other general circuits relied on controlled-not gate for entanglement. The chosen gate set for each circuit was selected to reflect common implementations and could be further improved to facilitate the overhead reduction. 

\noindent\textbf{Resource Estimation:} For resource estimation, we used Microsoft’s Azure Quantum Resource Estimator (AQRE) with server access. We assumed a superconducting qubit, with a gate speed of 50ns, readout latency of 100ns and gate and readout fidelity of $10^{-4}$~\cite{beverland2022assessing}. We use surface code as the choice of QEC and 15 to 1 distillation factories. These choices are available as models in AQRE.
We also experimented with trapped ion-like configurations. While it provides different physical error rates as well as the overall physical requirements, our experiments found that such a choice does not make much difference since the total number of cuts is too high to be compromised by the hardware choice. We will not discuss this in our main results.

\noindent\textbf{Metrics:} We assume that circuit cutting is implemented with one quantum computer and subcircuits are executed sequentially on that device. For the resource estimation on the reduction of number of physical qubits, we will use the metric, $\text{Percent Reduction} = 100\% \times ({N^{\text{phy}}}_{\text{baseline}} - \max \{N^{\text{phy}}_{c_1} \dots N^{\text{phy}}_{c_{N_{\text{pool}}}}\})/{N^{\text{phy}}}_{\text{baseline}}$, where $N^{\text{phy}}_{\text{baseline}}$ is the original circuit's physical qubits requirement and $N^{\text{phy}}_{c_i}, c_i \in C$ is the number of physical qubits count for each subexperiments $c_i$ in the pool $C$.
Other metrics that we are report are the number of cuts obtained after circuit cutting and the quantum runtime after circuit cutting computed with equation \ref{samplingcost}.  

\section{Results}    
\begin{table}
\caption{The quantum runtime is derived from resource estimation data. Values marked as `N/A' indicate that the runtime is too large to evaluate. The numbers following each algorithm name represent the problem size, measured in the number of logical qubits. Baseline runtime and physical qubits refers to metrics of the original full circuit using the resource estimator. The \textit{cutting runtime} represents the execution time for the subcircuits after partitioning the circuit. For this analysis, we only consider cuts that divide the circuit approximately in half by the number of qubits. The final column shows the reduction ratio of physical qubits after cutting. Note that the quantum runtime after partitioning provides only a lower bound. The factor $\varepsilon_{\text{rct}}$, which accounts for reconstruction error is not included in the calculations. Incorporating $\varepsilon_{\text{rct}}$ would further increase the estimated runtime.
} 
\centering
\small
\scalebox{1}{
\begin{tabular}{|c|c|c|c|c|}
\hline
Algorithm & \makecell{Baseline \\ runtime} (s) & \makecell{Cutting\\runtime} (s) & \makecell{Baseline \\ phy. qubits} & \makecell{\% reduction \\ phys. qubits}\\
\hline
QFT $10$         & 0.07             & 2070                & $10^{4.43}$ & 20\%      \\ \hline
QFT $20$         & 0.41             & 5493                & $10^{4.69}$ & 30\%      \\ \hline
QFT $30$         & 1.24             & $1.3\times 10^{4}$  & $10^{4.87}$ & 41\%      \\ \hline
QFT $40$         & 2.41             & $1.7\times 10^{4}$  & $10^{4.95}$ & 38\%      \\ \hline
QFT $50$         & 3.68             & $1.7\times 10^{4}$  & $10^{5.02}$ & 27\%      \\ \hline
QFT $60$         & 4.83             & $1.7\times 10^{4}$  & $10^{5.16}$ & 47\%      \\ \hline
Ising $3\times3$ & 20.64            & $1.0\times 10^{9}$  & $10^{4.53}$ & 45\%      \\ \hline
Ising $4\times4$ & 40.04            & $1.5\times 10^{9}$  & $10^{4.70}$ & 30\%      \\ \hline
Ising $5\times5$ & 69.55            & $1.6\times 10^{11}$ & $10^{4.88}$ & 22\%      \\ \hline
Ising $6\times6$ & 103.70           & $1.6\times 10^{11}$ & $10^{4.99}$ & 34\%      \\ \hline
HB $3\times3$    & 53.65            & $3.2\times 10^{11}$ & $10^{4.58}$ & 9\%       \\ \hline
HB $4\times4$    & 105.98           & $1.7\times 10^{4}$  & $10^{4.70}$ & 10\%      \\ \hline
HB $5\times5$    & 203.01           & $1.7\times 10^{4}$  & $10^{4.88}$ & 12\%      \\ \hline
HB $6\times6$    & 303.37           & $1.7\times 10^{4}$  & $10^{4.99}$ & 10\%      \\ \hline
FH $2\times2$    & 85.03            & N/A                 & $10^{4.43}$ & N/A       \\ \hline
FH $3\times3$    & 290.58           & N/A                 & $10^{4.64}$ & N/A       \\ \hline
FH $4\times4$    & 581.22           & N/A                 & $10^{4.70}$ & N/A       \\ \hline
QAOA $10$        & 0.15             & $8.0\times 10^{7}$  & $10^{4.43}$ & 20\%      \\ \hline
QAOA $20$        & 0.31             & $4.2\times 10^{31}$ & $10^{4.69}$ & 18\%      \\ \hline
QAOA $30$        & 0.52             & $3.4\times 10^{39}$ & $10^{4.80}$ & 20\%      \\ \hline
QAOA $40$        & 0.60             & $4.2\times 10^{31}$ & $10^{4.95}$ & 18\%      \\ \hline
QAOA $50$        & 1.05             & $1.7\times 10^{63}$ & $10^{5.02}$ & 18\%      \\ \hline
QAOA $60$        & 1.27             & $3.4\times 10^{39}$ & $10^{5.08}$ & 12\%      \\ \hline
QAOA $70$        & 1.50             & $4.2\times 10^{31}$ & $10^{5.14}$ & 16\%      \\ \hline
QAOA $80$        & 1.70             & $1.0\times 10^{39}$ & $10^{5.18}$ & 18\%      \\ \hline
QAOA $90$        & 1.94             & $1.8\times 10^{63}$ & $10^{5.23}$ & 20\%      \\ \hline
QAOA $100$       & 2.30             & $2.2\times 10^{55}$ & $10^{5.26}$ & 22\%      \\ \hline
QPE $2$          & 0.01             & $4.3\times 10^{7}$  & $10^{4.17}$ & 21\%      \\ \hline
QPE $4$          & 0.10             & $1.2\times 10^{61}$ & $10^{4.39}$ & 29\%      \\ \hline
QPE $6$          & 0.66             & N/A                 & $10^{4.58}$ & N/A       \\ \hline
QPE $8$          & 5.17             & N/A                 & $10^{4.70}$ & N/A       \\ \hline
Random $5$       & 0.02             & $8.9\times 10^{16}$ & $10^{4.20}$ & 13\%      \\ \hline
Random $10$      & 0.26             & N/A                 & $10^{4.49}$ & N/A       \\ \hline
Random $15$      & 1.00             & N/A                 & $10^{4.63}$ & N/A       \\ \hline
Random $20$      & 2.70             & N/A                 & $10^{4.76}$ & N/A       \\ \hline
BE $2$           & 0.01             & $6.8\times 10^{8}$  & $10^{3.94}$ & 8\%       \\ \hline
BE $3$           & 0.04             & $4.7\times 10^{17}$ & $10^{4.17}$ & 14\%      \\ \hline
BE $4$           & 0.21             & $1.4\times 10^{34}$ & $10^{4.35}$ & 8\%       \\ \hline
BE $5$           & 1.04             & $7.7\times 10^{65}$ & $10^{4.51}$ & 18\%      \\ \hline
\end{tabular}
}
\label{runtime_table}
\end{table}
\subsection{Number of cuts vs. physical qubit reduction}
\textbf{How to Interpret Figure~\ref{algo}:}  
The x-axis represents the parameter \textit{Max number of qubits per subcircuit}, which determines the maximum number of qubits that any subcircuit can contain during the cut-searching stage. For example, a value of 44\% (or 44) in a 9-qubit quantum algorithm implies that the maximum number of qubits per subcircuit is bounded by \( 9 \times 44\% \approx 4 \). For the QFT and QAOA benchmarks, this parameter is not shown. Instead, the x-axis displays the total number of qubits in the circuits. However, for these benchmarks, the maximum number of qubits corresponding to each bar increases at regular intervals. So, it should only involve simple calculation to get the qubit number. The y-axis represents either the number of cuts required for the circuits or the reduction in physical qubits, as defined in the experimental setup. Finally, the legend in each plot provides information about the problem size. Specifically, for block-encoding benchmarks, \( 2^N \) indicates that the goal is to encode a unitary of size \( 2^N \times 2^N \). For QPE benchmarks, the problem size corresponds to the size of the phase of the unitary being estimated.

Figure~\ref{algo} (i-p) shows the number of cuts required for each algorithm. These plots contain two variables, the problem size and the maximum number of qubits allowed in a subcircuit (maximal circuit width after applying cuts). There are three clear trends. First, as the problem size, that is, the number of qubits and circuit depth increases, the number of cuts increases. This is because larger circuits require more partitioning to obtain subcircuit qubit counts that are within the limits imposed by the maximum qubits per subcircuit parameter. Second, as the maximum qubits per subcircuit increase, the number of required cuts reduces because a small number of large subcirucits are sufficient to partition the original circuit. 

Third, the number of cuts is influenced by algorithm structure. Ising, Heisenberg and Fermi-Hubbard models typically require over a thousand cuts since they have nearest-neighbor gates and large depth. For these circuits, gate cuts are primarily used instead of wire cuts because of the highly entangling structure of the circuit. QFT and QPE circuits also require a lot of cuts since they have all-to-all gate patterns that are hard to cluster. Random circuits exhibit comparable behavior to Hamiltonian simulations, which is expected given their similar entangling structure.

Figure~\ref{algo} (a-h) shows the percentage reduction in the number of physical qubits needed after circuit cutting, compared to the baseline which doesn't use circuit cutting. The number of physical qubits reduces across algorithms and problem sizes when circuit cutting is used, with reduction ratios ranging from 5 to 80\%. Similar to the trends in number of cuts, the percentage reduction in physical qubit usage improves with problem size. This is because when large problems are broken down into very small subcircuit, we get significant reductions in the number of logical qubits, physical qubits and distillation resources required to implement it fault-tolerantly. When the maximum number of qubits per subcircuit is low, the physical qubit reductions are very high, but with increasing subcircuit size, the reductions dimish. When subcircuits are allowed by 80-90\% of the original problem size, circuit cutting gives approximately a 10\% qubit reduction.

\textit{This situation presents a trade-off: achieving a greater reduction in physical qubits necessitates more cuts, thus increasing the overhead, while fewer cuts allow a tolerable overhead at the cost of less reduction in qubit usage. This tradeoff is a challenge for circuit cutting as problem sizes, depth and complexity of gate structure increase.}

For instance, for an Ising model \(5 \times 5\), the subcircuits after cuts have at most 6 qubits per subcircuit at a data point of 24\%. The number of cuts required is around $10^{3.5}$ with a reduction of physical qubits around 60\%. While the number of cuts can be decreased by 25\% to separate the circuit with at most 11 qubits in each subcircuit. This manner decreases the physical qubits benefits by 10\%.

\subsection{Quantum and classical runtime overheads}
\textbf{Quantum runtime overhead:} Table~\ref{runtime_table} compares the quantum runtime overheads of circuit cutting executions with the baseline which does not use circuit cutting. Across benchmarks, quantum runtime is several orders of magnitude higher than the baseline.  As a comparison, the original circuit of a $6\times6$ Ising model only need 103 seconds for execution, but after cutting it requires 5000 years. For all benchmarks, except QFT, the runtime is extremely high, typically in the range of years, when circuit cutting is used. This is one of the main reasons why circuit cutting is infeasible in practice for fault-tolerant executions. This arises because of the number of cuts required to partition the algorithm and the types of gates used in the algorithm. For circuits like block encoding and random entangled circuits, the native gate is often the CNOT gate which has a large sampling overhead (Table \ref{sampling_overhead}). For QFT, the runtime is acceptable because QFT's gate type is the controlled phase gate with many small angle rotations which allow a small sampling overhead (CPhaseGate in Table \ref{sampling_overhead}). However, circuit cutting is still infeasible for QFT, as we discuss next.

\textbf{Classical post-processing overhead:} We consider the classical post-processing overhead, which scales as $O(4^{cuts})$ FLOPs on a classical computer. If we have 100 cuts, this would mean that $10^{60}$ FLOPs are approximately required, which is beyond the reach of any foreseeable classical supercomputer. Even in the case of QFT which has gate types that are suitable for cutting, the number of cuts is at least 100 for problem sizes beyond 50 qubits that are relevant for quantum advantage experiments. Therefore, across our applications, this exponential classical overhead makes circuit cutting infeasible for practical-scale fault-tolerant executions. 

The only benchmark which has acceptable classical overheads is QAOA, with the circuits feasible for cutting up to around 50 qubits. This is closely tied to how the algorithm’s circuit structure mirrors the topology of the problem graph. Since QAOA’s circuit  inherently matches the connectivity and interactions of the graph it operates on, this allow good partitioning if the graph can be clustered well. If the underlying problem graphs are more diverse, it is unclear if circuit cutting will still be feasible. Since QAOA's quantum overheads are very high, we did not explore this further.

\textit{Both quantum and classical overheads should be considered while evaluating techniques such as circuit cutting. For certain benchmarks we see that sometimes quantum or classical overheads with circuit cutting are acceptable, but none of our benchmarks show all desirable circuit cutting benefits of reductions in physical qubit usage, low quantum overhead and low classical overhead.}

\subsection{Theoretical analysis of number of cuts}
To validate the experimental estimates, we developed theoretical models to estimate cut counts for four benchmarks. \label{theoretical}

\noindent\textbf{Ising, Heisenberg and Fermi-Hubbard models:} These Hamiltonian simulations are implemented using a well-known algorithm called Trotterization,

an approximation technique that decomposes the time evolution operator \( e^{-iHt} \) of a Hamiltonian \( H \) into simpler, implementable steps. The Trotter formula approximates the evolution by splitting \( H \) into a sum of terms \( H = \sum_j H_j \), evolving each term individually, and repeating this sequence over multiple small time steps, $N_t$. However, the approximation introduces a \textit{Trotter error} (algorithmic error) that scales with the commutators of the Hamiltonian terms.  We use the fourth order Trotter approximation based on~\cite{beverland2022assessing}, where the error bound scales as 

\begin{equation}
    \epsilon_{\text{alg}} = \mathcal{O}(\frac{(t \sum_{j < k} \|[H_j, H_k]\|)^{1+p}}{N_t^p}),
    \label{ealgo}
\end{equation}
where $p$ is the order of the algorithm. With the scaling relation we can easily invert the expression to find the algorithmic error, depending on the number of trotter steps.

For a $D\times D$ square lattice, we have $D^2$ qubits in the grid for an Ising model. When we divide the $D\times D$ grid into subcircuits, each containing no more than $m$ qubits. The number of subcircuits, $N_{\text{subcircuits}}$ is approximately:

\begin{equation}
    N_{\text{subcircuits}} = \frac{D^2}{m}.
\end{equation}
The boundary of each subcircuit has qubits that interact with qubits in adjacent subcircuits. If a qubit is located near the boundary of a subcircuit, its interactions with neighbouring qubits in other subcircuits will need to be cut. The number of interactions that span across subcircuits depends on the size and shape of the subcircuits. Assuming a square lattice model, the number of boundary qubits scales roughly as $\sqrt{m}$
\begin{equation}
    \text{Boundary qubits} \approx  \frac{D^2}{m} \cdot \sqrt{m}.
\end{equation}
Therefore, the minimum number of cuts needed to divide the 2D Heisenberg model circuit into subcircuits is:
\begin{equation}
    \text{Number of cuts} \approx 4 \cdot \text{number of trotter steps} \cdot \frac{D^2}{\sqrt{m}}.
    \label{cut hb}
\end{equation}
However, for the Heisenberg's model, each qubit interacts with its nearest neighbours through 3 types of controlled-phase gate, $R_{zz}(\theta), R_{xx}(\theta), R_{yy}(\theta)$, in the Trotterized evolution step. So, the constant scaling factor is 12 for the number of cuts comparing to 3 for in Ising model, where only $R_zz$ interation is considered. For the Fermi-Hubbard model, the scaling changes again because of the incorporation of both the spin-up and spin-down states, resulting in a total number of qubits of $2\times D \times D$ in a $D \times D$ lattice.

Our aim to was scale up problems like Ising up to $10\times 10$ --- practical sizes, but we found that cutting tools failed beyond $6 \times 6$. For example, in Figure~\ref{single}, the number of cuts is calculated for each lattice size up to $10 \times 10$, considering a \textbf{single} Trotter step. The results suggest that post-processing might be feasible due to the manageable number of cuts observed in this case, which aligns with the approach commonly taken by researchers in the literature. However, when scaling to practical scenarios involving many Trotter steps, the number of cuts increases linearly with the number of steps, rendering the problem impractical. A preliminary analysis shows that the average number of cuts across all percentages of subcircuit qubit classifications ranges from $10^{3.2}$ to $10^{4.1}$. Experimental results are consistent with this estimate, yielding a range of values from  $10^{3.3}$ to $10^{3.8}$. For the $6 \times 6$ lattice model, the predictions from Eqn.~\eqref{cut hb} are confirmed: as $m$ increases, the rate of reduction in the number of cuts gradually decreases. 

\noindent\textbf{QFT:} QFT has a structure where each qubit has controlled phase gates with all preceding qubits. By partitioning the circuit into two parts, we can estimate the number of cuts required. For each pair of adjacent subcircuits, there will be controlled-phase gates that need to be cut. The number of such cuts can thus be approximated by:
\begin{equation}
    \text{Number of cuts} \approx \sum^{\lceil \frac{n}{m} \rceil - 1}_{k=1} (m \cdot (n-km)),
    \label{qfteqn}
\end{equation}
where $n$ is the number of qubits in the original circuit and $m$ is the size of one of the subcircuits. Intuitively, as the number of cuts increases as $n$ increases for a fixed $m$ since more qubits need to be distributed across subcircuits, leading to more connections between qubits in different subcircuits.
As the parameter $m$ increases, fewer cuts are required to partition the circuit, because more qubits can remain within a single subcircuit. Experimental data demonstrate an average increase in the number of cuts from $10^{0.75}$ to $10^{3.5}$ as the number of qubits increases from 5 to 60. This trend aligns closely with the approximation from Eqn.~\eqref{qfteqn}, which predicts values ranging from $10^{0.84}$ to $10^{3.5}$, with minimal deviation from the observed results.

\emph{These theoretical models support our experimental findings by showing that the number of cuts scales with problem size, which translates to increasing exponential classical overheads as we scale up. } 

These theoretical models are derived from ``natural'' cuts of an algorithm. For example, in the case of Hamiltonian simulations, the cuts we consider above are the natural partitions of a lattice into smaller lattices. We do not prove that they are theoretically optimal, but we do not expect significant reductions in cut counts for these algorithms, even with advanced graph partitioning strategies. Similarly, GPU acceleration or tensor contraction optimizations such as \cite{tornow2024quantum} are effective for small problem sizes, but cannot cope with the exponential compute growth as problem sizes increase.

\section{Related Work}
\textbf{Types of cut:} A variety of circuit cutting techniques have been developed in addition to the qubit and gate cuts~\cite{harada2023doubly, ufrecht2023cutting}. These include cutting techniques that are dependent on the type of gate that is cut~\cite{qiskit-addon-cutting} and the ability to use classical communication to reduce overhead.~\cite{harada2023doubly, Brenner2023optimal, Lowe2023fastcutting, piveteau2022circuit} assume classical communication and \cite{peng2020simulating, tang2021cutqc, mitarai2021constructing} assume there is no classical communication . Local operations (LO) refer to the scenario where the two computers can only perform operations locally and operate entirely independently. In contrast, Local Operations and Classical Communication (LOCC) allow for two-way classical communication between the computers. Utilizing LOCC can optimize overhead, particularly with improved scaling for multiple cut wires. However, it offers no advantage when quasi-probabilistically simulating a single instance of a gate~\cite{piveteau2022circuit}. The benefits of LOCC become apparent only when multiple gates are cut simultaneously, where the overhead is reduced. Additionally, classical communication requires the preparation of ancillas in Bell states, introducing the challenge of additional resource requirements. A critical consideration is whether these extra ancillas remain idle. If so, additional QEC qubits are necessary to maintain fault tolerance. Consequently, for the scope of our current work, we focus exclusively on the ancilla-free method.

\noindent\textbf{Optimization strategies:} Several works have explored how to mitigate the exponential increase in the number of required measurements, with respect to the number of cut locations. Recent strategies include using randomized measurements~\cite{Lowe2023fastcutting, chen2022quantum}, identifying optimal cut points through golden circuit cutting techniques~\cite{chen2023efficient, chen2023online}, and applying variational optimization~\cite{Uchehara2022}. For the LOCC settings, further studies~\cite{Lowe2023fastcutting, pednault2023alternative, harada2023doubly} have reduced the required number of ancilla qubits by using either a random measurement basis or optimized decomposition techniques. 

\noindent\textbf{Tensor network methods:} Circuit cutting methods are closely related to tensor networks (TN).TN represent quantum states in a compressed form and provide a framework for characterizing entanglement patterns. Prior research~\cite{tornow2024quantum} introduced the concept of hybrid TN to reduce the overhead of classical post-processing. Adaptive circuit cutting (ACK)~\cite{Jmartin} demonstrates improvements in the overhead by representing circuits in linear depth~\cite{lin2021real} with TN.

\section{Conclusions}
Circuit cutting is a technique for orchestrating the execution of large quantum programs on small quantum computers. There has been a lot of research interest in this technique in the last few years, with several lines of theoretical work to reduce the overheads, experimental implementations that showcase the technique on current QC systems and the development of automated cut finding frameworks. Since this technique is seen as a promising way of using future distributed quantum hardware~\cite{IBM}, our work evaluates whether cutting techniques can indeed scale to the needs of practical applications running on fault-tolerant hardware. Unfortunately, we find that the classical and quantum runtime overheads of circuit cutting are too high for practically-relevant benchmarks. In particular, applications of QC such as Hamiltonian simulation where we expect the first practically-useful quantum speedups over classical computing are especially problematic for circuit cutting due to their gate structures.

Our work aligns with a recent theoretical argument~\cite{jing2024circuit} that shows that the sampling overhead of circuit-knitting is exponentially lower bounded by the entanglement cost. 

Circuit cutting remains a valuable tool for small-scale quantum algorithms, short depth circuits or Hamiltonian simulations with short evolution times. In this setting, further research is needed to reduce classical post-processing costs, such as choosing the optimal gate set for cutting, and mitigate the exponential scaling limitations identified in both practical benchmarks and theoretical analysis. If applications are co-designed with circuit cutting, with highly clustered gate structures that match the requirements of cutting, circuit cutting could be scaled up in a resource-efficient manner. 

\bibliographystyle{ACM-Reference-Format}
\bibliography{sample-base}

\end{document}